\documentclass[aps,prl,twocolumn,showpacs,superscriptaddress,superscriptaddress,amsthm,amsmath,amssymb]{revtex4-1}  
\usepackage{graphicx}
\usepackage{dcolumn}
\usepackage{bm}
\usepackage{hyperref}
\usepackage{mathtools}
\usepackage{color}
\usepackage{mhchem}
\usepackage{float}
\usepackage{bbm}
\usepackage{pgfplots}
\usepackage{tikz-qtree}

\usepackage{amssymb,amsmath,amsthm,amsfonts}

\input{Qcircuit}

\begin{document}

\footnotetext{This manuscript has been authored by UT-Battelle, LLC, under Contract No. DE-AC0500OR22725 with the U.S. Department of Energy. The United States Government retains and the publisher, by accepting the article for publication, acknowledges that the United States Government retains a non-exclusive, paid-up, irrevocable, world-wide license to publish or reproduce the published form of this manuscript, or allow others to do so, for the United States Government purposes. The Department of Energy will provide public access to these results of federally sponsored research in accordance with the DOE Public Access Plan.}

\newcommand{\comment}[1]{\color{black} \color{black}}

\title{Toward convergence of effective field theory simulations on digital quantum computers}

\author{O.~Shehab}
\affiliation{IonQ, Inc, 4505 Campus Drive
College Park, MD 20740, USA}

\author{K. ~Landsman}
\affiliation{Joint Quantum Institute, Department of Physics and Joint Center for Quantum Information and Computer Science, University of Maryland, College Park, MD 20742}

\author{Y. ~Nam}
\affiliation{IonQ, Inc, 4505 Campus Drive
College Park, MD 20740, USA}

\author{D. ~Zhu}
\affiliation{Joint Quantum Institute, Department of Physics and Joint Center for Quantum Information and Computer Science, University of Maryland, College Park, MD 20742}

\author{N. M. ~Linke}
\affiliation{Joint Quantum Institute, Department of Physics and Joint Center for Quantum Information and Computer Science, University of Maryland, College Park, MD 20742}

\author{M. ~Keesan}
\affiliation{IonQ, Inc, 4505 Campus Drive
College Park, MD 20740, USA}


  




\author{R. C. Pooser}
\affiliation{Computational Sciences and Engineering Division, Oak Ridge National Laboratory,
Oak Ridge, TN 37831, USA}
\affiliation{Department of Physics and Astronomy, University of Tennessee,
Knoxville, TN 37996, USA} 



\author{C. ~Monroe}
\affiliation{Joint Quantum Institute, Department of Physics and Joint Center for Quantum Information and Computer Science, University of Maryland, College Park, MD 20742}
\affiliation{IonQ, Inc, 4505 Campus Drive
College Park, MD 20740, USA}

\date{\today}

\begin{abstract}
We report results for simulating an effective field theory to compute the binding energy of the deuteron nucleus using a hybrid algorithm on a trapped-ion quantum computer.
Two increasingly complex unitary coupled-cluster ansaetze have been used to compute the binding energy to within a few percent for successively more complex Hamiltonians.
By increasing the complexity of the Hamiltonian, allowing more terms in the effective field theory expansion and calculating their expectation values, we present a benchmark for quantum computers based on their ability to scalably calculate the effective field theory with increasing accuracy. Our  result of $E_4=-2.220 \pm 0.179$MeV may be compared with the exact Deuteron ground-state energy $-2.224$MeV. We also demonstrate an error mitigation technique using Richardson extrapolation on ion traps for the first time. The error mitigation circuit represents a record for deepest quantum circuit on a trapped-ion quantum computer.
\end{abstract}

\maketitle

\section{Introduction}

Simulating Fermonic matter using quantum computers has recently become an active field of research. With the advent of noisy intermediate-scale quantum (NISQ) devices that are capable of processing quantum information, hybrid quantum-classical computing (HQCC) has been proposed to be a worthy strategy to harness the advantage quantum computers provide as early as possible. A host of HQCC demonstrations, ranging from its application in chemistry \cite{omalley2016, kandala2017, nam2019ground} to machine learning \cite{QBAS}, are in fact already available in the literature. 

NISQ devices are however susceptible to errors and defects. Thus, the quantum circuits to be run on these machines need to be sufficiently small so that the results that the quantum computers output are still useful. On the other hand, in order for the quantum computational results to be useful, the computation that the quantum computer performs needs to be sufficiently demanding such that readily available classical devices cannot easily arrive at the same results. However, there is a lack of empirical evidence for the performance scaling of HQCC as problems become more complex. A test, or benchmark, of this scalability would be useful to inform future quantum algorithm development.

Here, using the effective field theory (EFT) simulation of a deuteron, we outline a path to scalable HQCC and provide a benchmark that determines the HQCC performance scaling of a quantum computer.  We further demonstrate that a trapped-ion quantum computer today is capable of addressing small, yet scalable HQCC problems, and that it shows promises toward scaling to reliable computational results when a quantum advantage is demonstrated.

We also demonstrate a re-parametrization technique that yields a quantum circuit amenable to implementation on quantum computers with nearest-neighbor connectivity. We report our experimental results that leverage known error mitigation techniques \cite{temme2017error, li2017efficient, endo2018practical,mcardle2018error}. The theoretical predictions for the three- and four-qubit case are within the error bars of the experimental results.

\section{Hamiltonian and Ans\"atz} 

The $N$ oscillator-basis deuteron Hamiltonian we consider (see Supplementary material for detail) is
\begin{equation}
\label{HN}
H_N = \sum_{n,n'=0}^{N-1} \langle n'|(T+V)|n\rangle a^\dagger_{n'} a_{n},
\end{equation}
where the operators $a^\dagger_{n}$ and  $a_{n}$ create and annihilate a deuteron in the harmonic-oscillator $s$-wave state $|n\rangle$ and the matrix elements of the kinetic and potential energy are 
\begin{eqnarray}
\langle n'|T|n\rangle &=& {\hbar\omega\over 2}\bigg[ (2n+3/2)\delta^{n'}_n - \sqrt{n(n+1/2)}\delta_n^{n'+1}\nonumber\\
&&- \sqrt{(n+1)(n+3/2)}\delta_n^{n'-1}\bigg] , \nonumber\\
\langle n'|V|n\rangle &=& V_0 \delta_{n}^0\delta^{n'}_n,
\end{eqnarray}
where $\hbar\omega \approx 7\,{\rm MeV}$ and $V_0 \approx -5.68\, {\rm MeV}$. Since our goal is to find the ground state energy expectation values as a function of $N$ using a quantum computer, we apply Jordan-Wigner transform \cite{jordan1993paulische} to our physical Hamiltonian in (\ref{HN}) to find the qubit Hamiltonian. For $N=2,3,$ and $4$, we have
\begin{eqnarray}
\label{QH}
H_2 &=& 5.907 I + 0.218 Z_0 -6.125 Z_1 - 2.143  (X_0X_1 + Y_0Y_1) \nonumber \\
H_3 &=& H_2 + 9.625 (I - Z_2) -3.913(X_1X_2+Y_1Y_2) \nonumber \\
H_4 &=& H_3 + 13.125 (I - Z_3) - 5.671 (X_2 X_3 + Y_2 Y_3). 
\end{eqnarray}

For our current example of a Deuteron EFT simulation, the UV cutoff determines the largest matrix element in the nuclear Hamiltonian, which controls the scaling of the coefficients of the Pauli terms in the qubit Hamiltonian in (\ref{QH}). Since the uncertainty in determining the expectation value of the Hamiltonian is bounded by the largest absolute value of the coefficients in the qubit Hamiltonian \cite{kandala2017}, the higher the UV cutoff, the larger the uncertainty in the expectation value of the Hamiltonian becomes. To meet the required, preset uncertainty, we need to make a larger number of measurements for a large-coefficient Hamiltonian.
Because the largest coefficient tends to grow with basis size, this effectively induces an implementation-level tug-of-war between the increasingly accurate simulation from considering a larger oscillator basis and the accumulation of errors on NISQ devices susceptible to, e.g., drifts, that occur over the required, longer overall runtime. While frequently calibrating the quantum computer may help reduce the errors, this may not be desirable as it would significantly increase the resource overhead.

For the HQCC ansatz, we use the $N$-site unitary coupled-cluster singles (UCCS) ansatz
\begin{equation}
\label{UCCS}
|\Psi_{\text{UCCS}}\rangle = \exp\left(\sum\limits_{k=1}^{N-1}\theta_{k} [a^{\dagger}_{0}a_{k} - a^{\dagger}_{k}a_{0}]\right)|1_0\rangle,
\end{equation}
where $\vec{\theta} = \{\theta_1,...,\theta_{N-1}\}$ is the set of $N-1$ real-valued variational parameters and $|1_i\rangle$ denotes the state $|0,...,0,1,0,...,0\rangle$ with the $i$th $s$-wave state occupied. We compute the deuteron binding energy by minimizing the quantum functional $\langle\Psi_\text{UCCS}|H_{N}|\Psi_{\text{UCCS}}\rangle$ with respect to $\vec{\theta}$. The initial state $|1_0\rangle = |1,0,0,...0\rangle$ represents the occupation of the 0th $s$-wave state.

To implement the UCCS ansatz on our quantum computer, we re-parameterized (\ref{UCCS}) in the hyper-spherical coordinate, i.e.,
\begin{eqnarray}
\label{Eq:UCCWF}
|\Psi_{\text{UCC}}\rangle & = & \sum\limits_{k=0}^{N-2}\cos(\lambda_{k})\tilde{|1_{k}\rangle} + \tilde{|1_{N-1}\rangle},
\end{eqnarray}
where $\tilde{|1_k\rangle} \equiv \prod\limits_{i=0}^{k-1}\sin(\lambda_{i})|1_{k}\rangle$ with $\tilde{|1_{0}\rangle}=|1_{0}\rangle$. This choice is deliberate and exact, since the excitation operator in (\ref{UCCS}) is solely composed of single-excitations. Note we have relabeled (and re-indexed) variational parameters as $\vec{\lambda} = \{\lambda_0,...,\lambda_{N-2}\}$.

With the new parameterization shown in (\ref{Eq:UCCWF}), we may now synthesize the ansatz circuit straightforwardly. Let us define the amplitude shifting unitary $U_{i,i+1} (\lambda_i) \equiv (C_{i+1}X_{i}) (C_iRY_{i+1}(\lambda_i))$, where $C_{m}G_{n}$, for instance, denotes a single-qubit gate $G$ acting on qubit $n$, controlled by qubit $m$, such that $U(\lambda) (\alpha|00\rangle + \beta |10\rangle) = \alpha|00\rangle + \beta (\cos{\lambda}|10\rangle + \sin{\lambda}|01\rangle$). Applying $U_{i,i+1}$ in series to an initial state of $|1_0\rangle$, we have 
\begin{equation}
\label{Eq:UCC_hopping}
|\Psi_{\text{UCC}}\rangle = \left[\prod_{i=0}^{N-2} U_{i,i+1} (\lambda_i) \right]|1_0\rangle.
\end{equation}

For the first non-trivial case of $N=2$, we need to optimize $U_{0,1}$ acting on $|1_0\rangle$. Since the initial state is $|10\rangle$, $C_{0}iRY_{1} |10\rangle = iRY_{1} |10\rangle$. The optimized circuit $\mathcal{C}_{2}$ is
\[
\Qcircuit @C=1em @R=1.2em @!R{
     & |1\rangle & & \multigate{1}{\mathcal{C}_{2}} & \qw &  \\
     & |0\rangle & & \ghost{\mathcal{C}_{2}}           & \qw &  
     }
\raisebox{-1.15em}{\hspace{2mm}=\hspace{6mm}}
\Qcircuit @C=1em @R=.7em @!R{
     & |1\rangle & & \qw                                          & \targ     & \qw \\
     & |0\rangle & & \gate{{\rm e}^{i\lambda_0 Y}} & \ctrl{-1} & \qw
     }
\raisebox{-1.15em}{\hspace{2mm}.\hspace{6mm}}
\]
For $N>2$, we iteratively construct the circuit $\mathcal{C}_N$ as shown below.
\[
\Qcircuit @C=1em @R=.80em {
     & |1\rangle & & \multigate{3}{\mathcal{C}_{N}} & \qw \\
     & |0\rangle & & \ghost{\mathcal{C}_{N}}& \qw &  \\
     &\vdots&  & \ghost{\mathcal{C}_{N}} & \qw &  \\
     & |0\rangle & & \ghost{\mathcal{C}_{N}}& \qw &  
     }
\raisebox{-2.55em}{\hspace{2mm}=\hspace{6mm}}
\Qcircuit @C=1em @R=.7em {
      |1\rangle & & \multigate{2}{\mathcal{C}_{N-1}} & \qw & \qw & \qw & \qw\\
      |0\rangle & & \ghost{\mathcal{C}_{N-1}}& \qw & \qw &  \qw & \qw\\
      \vdots & & \ghost{\mathcal{C}_{N-1}}& \qw & \ctrl{+1}\ &  \targ & \qw \\
      |0\rangle & & \qw & \qw & \gate{{\rm e}^{i\lambda_{N-2} Y}} & \ctrl{-1} & \qw 
}
\]

\section{Results}
\label{Results}
We implemented our EFT simulation on an ion-trap quantum computer that may selectively load either five or seven \ce{^{171}Yb+} qubits. The qubit states $|0\rangle = |0, 0\rangle$ and $|1\rangle = |1, 0\rangle$ (with quantum numbers $|F, m_F\rangle$) are chosen from the hyperfine-split \ce{^{2}S_{1/2}} ground level with an energy difference of $12.64$ GHz. The $T_2$ coherence time with idle qubits is measured to be 1.5(5) sec, limited by residual magnetic field noise. The ions are initialized by an optical pumping scheme and are collectively read out  using state-dependent fluorescence detection \cite{olmschenk2007manipulation}, with each ion being mapped to a distinct photomultiplier tube (PMT) channel. State detection and measurement (SPAM) errors are characterized and corrected for in detail by inferring the state-to-state error matrix \cite{burrell2010high}. 

For the details of the single and two qubit gate implementations we refer the readers to Appendix A of \cite{landsman2018verified} and to \cite{debnath2016demonstration, molmer1999multiparticle, zhu2006arbitrary, choi2014optimal}. For the three qubit ansatz, we load five ions in the trap and use every other ion as qubit. For the four qubit ansatz, we load seven ions in the trap, using the inner 5 as qubits, with the outermost pair being used to evenly space the middle five ions. Entangling gates are derived from normal motional modes that result from the Coulomb interaction between ions, and the trapping potential. Off-resonantly driving both red and blue motional modes simultaneously leads to an entangling M{\o}lmer-S{\o}rensen interaction \cite{molmer1999multiparticle}.(See Supplementary material for circuits optimized for the native gate set.)

Logical qubits $0,1,2$, that denote $s$-wave states, are mapped to physical qubits $3,1,5$ in the three qubit experiment. The single qubit rotation fidelities are ${\sim} 99.5\%$ for each each ion. The XX gate fidelity \cite{kim2009entanglement, debnath2016programmable} is $99.3\%$, $97.7\%$, and $99.0\%$ on ion pairs $(1, 3)$, $(1, 5)$, and $(3, 5)$ respectively. The average 3-qubit state readout fidelity is $0.978$. For $H_4$ we map logical $1,2,3,4$ onto ions $1,2,3,5$. The measurement of even parity population for a maximally entangled XX gate are $99.4\%$, $99.8\%$, and $99.7\%$ for ions $(1, 2)$, $(2, 3)$, and $(3, 5)$. The averages of four qubit readout fidelity is $96.3\%$.

\begin{figure}[H]
\centering
\includegraphics[scale=0.40]{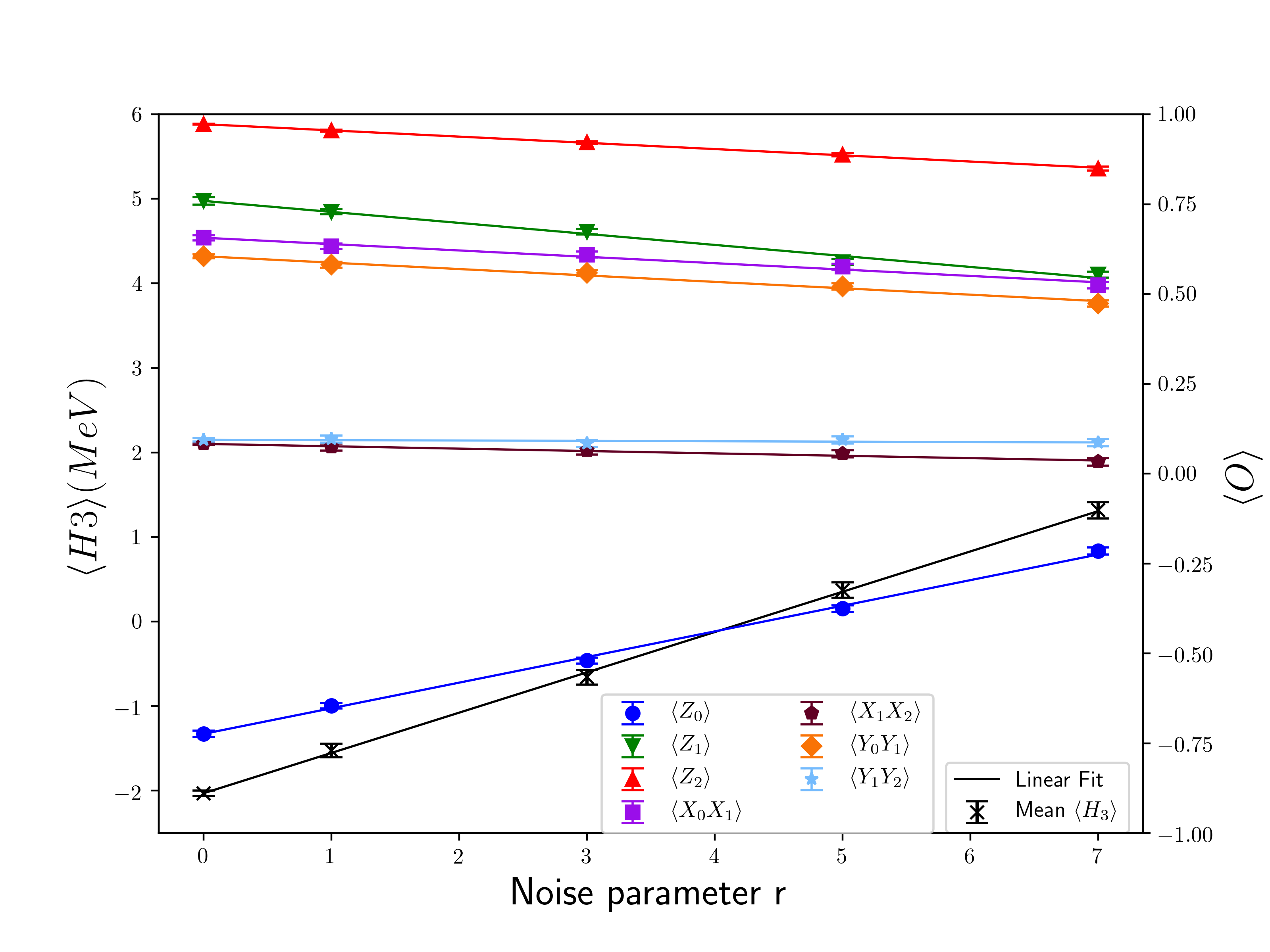}
\caption{Expectation values of Hamiltonian terms in $H_3$ as a function of noise parameter $r = 2M + 1$. Various colored, solid symbols are the expectation values of individual terms in $H_3$. Black crosses are $H_3$, computed according to Eq. (\ref{QH}). Colored, solid lines are the linear fits to the corresponding individual Hamiltonian terms in $H_3$. The black solid line is the linear fit to $H_3$. We use the linear fits to extrapolate to the zero noise limit. The error bars in the figure are statistical errors based on finite sampling and a binomial distribution. The binding energy is determined as $-2.030 \pm 0.034$MeV}
\label{fig:plot}
\end{figure}

Figure~\ref{fig:plot} shows the experimentally determined expectation value of the Hamiltonian $H_3$ at the theoretically predicted minimum $\lambda_0 = 0.250$ and $\lambda_1 = 0.830$. We employed the error minimization technique \cite{endo2018practical,mcardle2018error}, based on Richardson extrapolation \cite{richardson1927viii}, to our circuit by replacing all occurrences of $\text{XX}(\theta)$ with $\text{XX}(\theta)\mathbbm{1}^{M}$, where $\mathbbm{1}^M = [\text{XX}(-\theta)\text{XX}(\theta)]^M$ for $M=0,1,2,3$. The linearly-extrapolated, zero-noise limit shows $\langle H_3 \rangle = -2.030 \pm 0.034$MeV, which is in excellent agreement with the theoretically expected value of -2.046MeV.

\begin{figure}[H]
\centering
\includegraphics[scale=0.40]{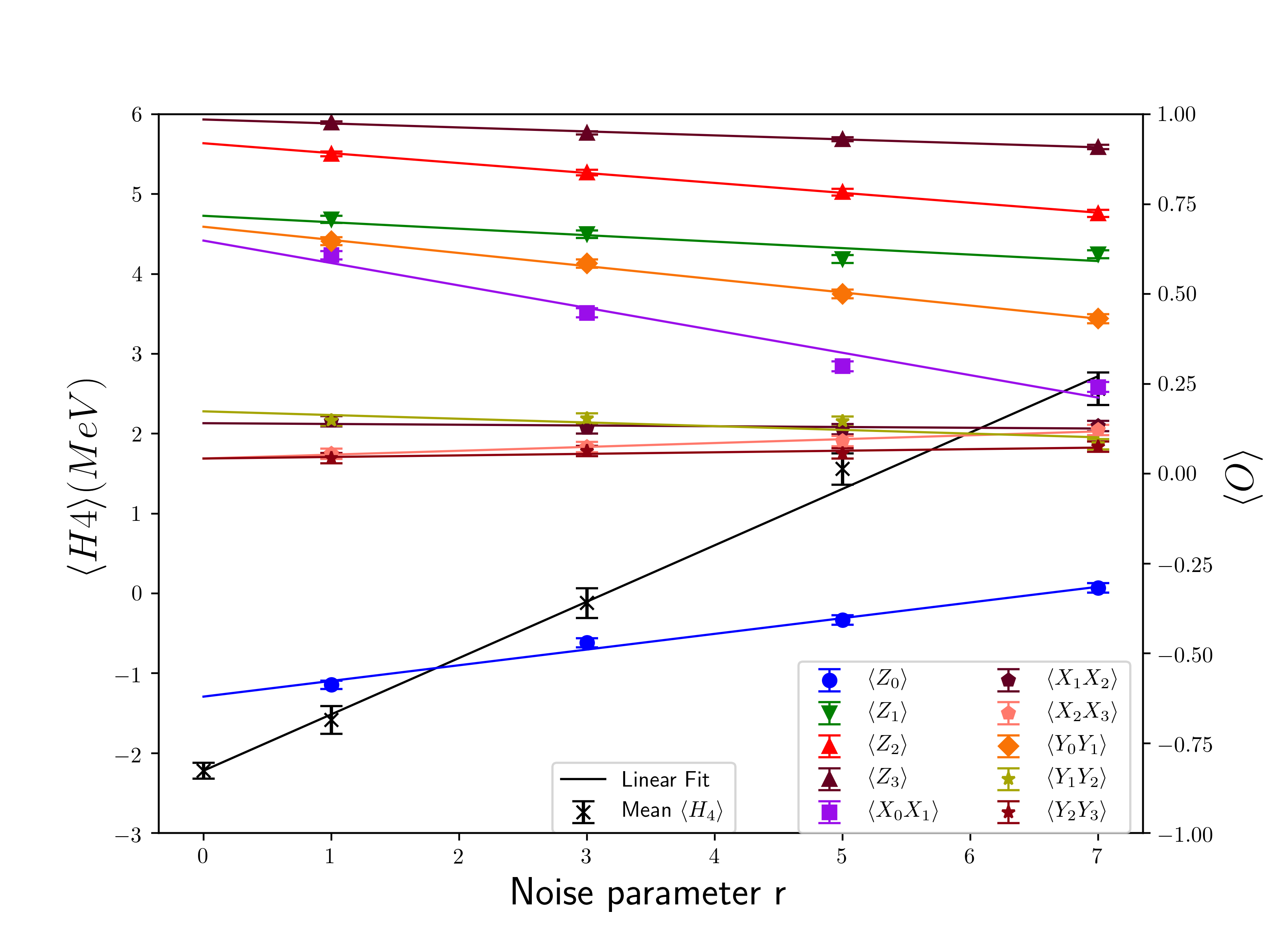}
\caption{Expectation values of Hamiltonian terms in $H_4$ as a function of noise parameter $r = 2M + 1$. Various colored, solid symbols are the expectation values of individual terms in $H_4$. Black crosses are $H_4$, computed according to Eq. (\ref{QH}). Colored, solid lines are the linear fits to the corresponding individual Hamiltonian terms in $H_4$. The black solid line is the linear fit to $H_4$. We use the linear fits to extrapolate to the zero noise limit. The error bars in the figure are statistical errors based on finite sampling and a binomial distribution. The binding energy is determined as $-2.220 \pm 0.179$MeV.}
\label{fig:plot4}
\end{figure}

Figure~\ref{fig:plot4} shows the analogous figure for $H_4$ evaluated at the theoretically optimal parameters $\lambda_0=0.8584$, $\lambda_1=0.9584$, and $\lambda_2 =0.7584$. The linearly-extrapolated, zero-noise limit shows $\langle H_4 \rangle = -2.220 \pm 0.179$MeV, again statistically consistent with the theoretically expected value of -2.143MeV. We note that the largest circuit that was run on our quantum computer to generate Figure~\ref{fig:plot4} involved implementing 35 two-qubit XX gates.

To further corroborate the accuracy of our quantum computational results, we also investigated the energy expectation values at various locations in the ansatz parameter space. Specifically, we explored the four-qubit ansatz's parameter settings that theoretically result in approximately $10\%$ or $20\%$ deviation from the theoretical minimum by varying one parameter at a time while fixing the other two constant to their optimal values. Table~\ref{tab:param} shows the choice of parameters and their respective, experimentally-obtained zero-noise-limit expectation values of $H_4$, compared with the theoretical values. We 

\begin{figure}
\centering
\includegraphics[width=\columnwidth]{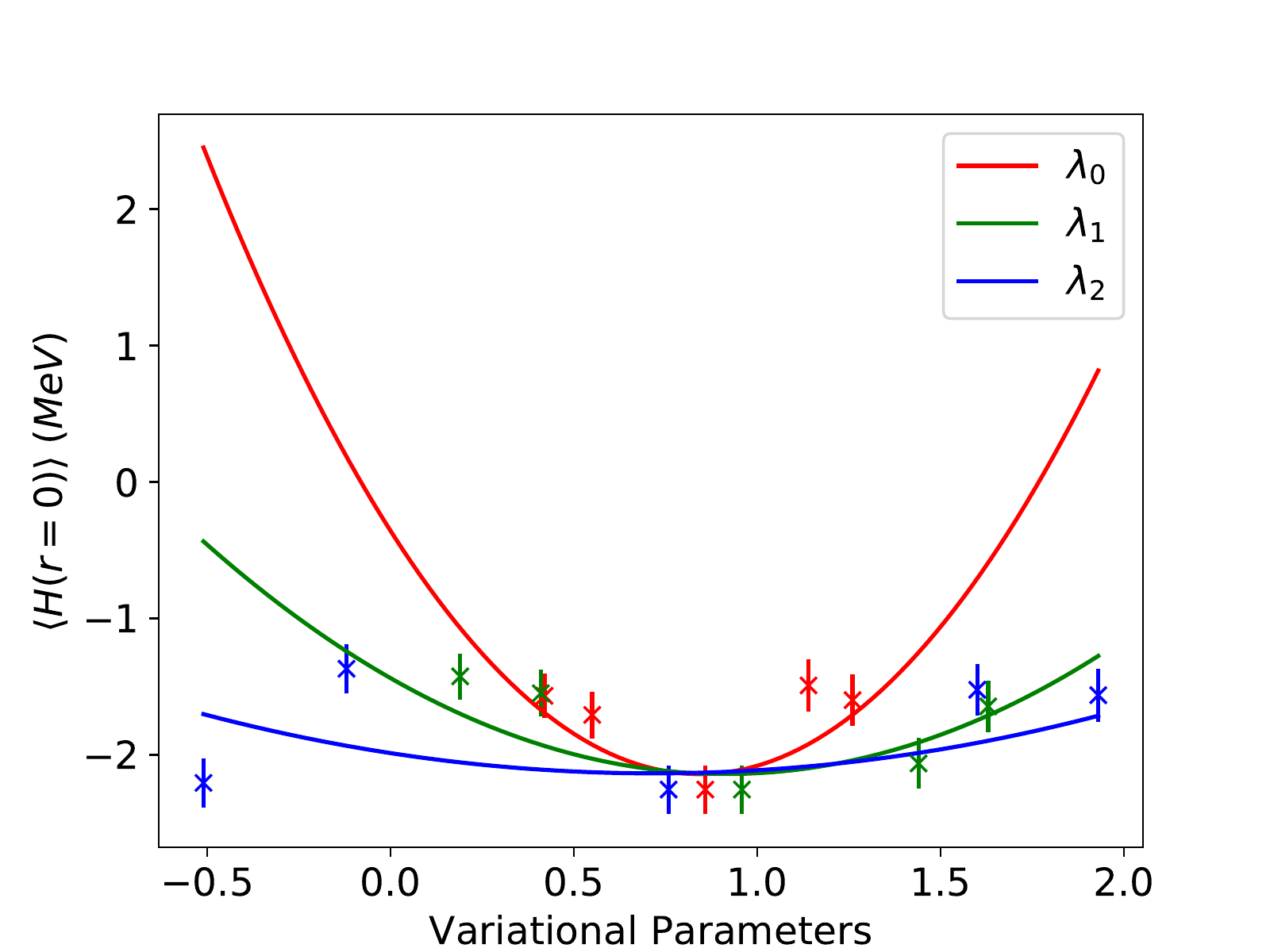}
\caption{Expectation value $\langle H_4 \rangle (r=0)$ as a function of a parameter chosen from the set $\{\lambda_0, \lambda_1, \lambda_2\}$. The plot symbols denote the zero-noise-limit extrapolated, also given in Table~\ref{tab:param}, and the solid lines denote the theoretical values.}
\label{fig:plot4basin}
\end{figure}

show in Fig.~\ref{fig:plot4basin} the data reported in Table~\ref{tab:param} as a visual aid. The minimal binding energy can be estimated by fitting each set of data to a quadratic form and minimizing the fit. Doing so results in individual estimates of $E_i = -2.080\pm-0.151$, $-2.200\pm0.149$, $-1.946\pm0.124$, for the three respective lambda parameters, with an average minima of $E=-2.088$ with $2.9\%$ error. Our computations therefore match previous error rates while increasing the system size, thus continuing to provide a path towards scalable simulations.

To further corroborate the accuracy of our quantum computational results, we also investigated the energy expectation values at various locations in the ansatz parameter space. Specifically, we explored the four-qubit ansatz's parameter settings that theoretically result in approximately $10\%$ or $20\%$ deviation from the theoretical minimum by varying one parameter at a time while fixing the other two constant to their optimal values. Table~\ref{tab:param} shows the choice of parameters and their respective, experimentally-obtained zero-noise-limit expectation values of $H_4$, compared with the theoretical values. We show in Fig.~\ref{fig:plot4basin} the data reported in Table~\ref{tab:param} as a visual aid. The minimal binding energy can be estimated by fitting each set of data to a quadratic form and minimizing the fit. Doing so results in individual estimates of $E_i = -2.080\pm-0.151$, $-2.200\pm0.149$, $-1.946\pm0.124$, for the three respective lambda parameters, with an average minima of $E=-2.088$ with $2.9\%$ error. Our computations therefore match previous error rates while increasing the system size, thus continuing to provide a path towards scalable simulations.


\begin{table}
\centering
\begin{tabular}{|c|c|c|c|c|}
\hline
$\lambda_0$ & $\lambda_1$ & $\lambda_2$ & $\langle H_4 \rangle [\text{experiment}]$ & $\langle H_4 \rangle [\text{theory}]$ \\
\hline
$0.858$ & $0.958$ & $0.758$ & $-2.256 \pm 0.179$ & $-2.143$ \\
\hline
$0.420$ & $0.958$ & $0.758$ & $-1.568 \pm 0.165$ & $-1.693$ \\
$0.550$ & $0.958$ & $0.758$ & $-1.708 \pm 0.172$ & $-1.925$ \\
$1.140$ & $0.958$ & $0.758$ & $-1.492 \pm 0.190$ & $-1.921$ \\
$1.260$ & $0.958$ & $0.758$ & $-1.599 \pm 0.191$ & $-1.708$ \\
\hline
$0.858$ & $0.190$ & $0.758$ & $-1.425 \pm 0.169$ & $-1.707$ \\
$0.858$ & $0.410$ & $0.758$ & $-1.549 \pm 0.172$ & $-1.916$ \\
$0.858$ & $1.440$ & $0.758$ & $-2.064 \pm 0.187$ & $-1.915$ \\
$0.858$ & $1.630$ & $0.758$ & $-1.646 \pm 0.188$ & $-1.707$ \\
\hline
$0.858$ & $0.958$ & $-0.510$ & $-2.066 \pm 0.179$ & $-1.713$ \\
$0.858$ & $0.958$ & $-0.120$ & $-1.370 \pm 0.182$ & $-1.917$ \\
$0.858$ & $0.958$ & $1.600$  & $-1.524 \pm 0.187$ & $-1.918$ \\
$0.858$ & $0.958$ & $1.930$  & $-1.563 \pm 0.194$ & $-1.709$ \\
\hline
\end{tabular}
\caption{Expectation value $\langle H_4 \rangle$ for various sets of variational parameters. $\langle H_4 \rangle [\text{experiment}]$ denote the zero-noise limit extrapolated values of $\langle H_4 \rangle$ obtained from our trapped-ion quantum computer. $\langle H_4 \rangle [\text{theory}]$ denote the corresponding, theoretically predicted values. All energies are measured in MeV. The top row shows the exact minimum configuration and results. The next set of four rows show the cases where we vary $\lambda_0$. The following two sets of four rows show the corresponding configuration-results pair for varying $\lambda_1$ and $\lambda_2$, respectively.}
\label{tab:param}
\end{table}

\comment{
Table~\ref{tab:basin} shows the experimental results employing the aforementioned minimization technique. We plot the zero-noise expectation values and compare them to the theoretical values in Fig.~\ref{fig:plot4basin}. 
}

\comment{
\begin{table}[H]
\centering
\scriptsize
\begin{tabular}{| c | c | c | c | c | c | c | c |}
\hline
$\lambda_0$& $\lambda_1$ & $\lambda_2$  & $\langle H_3 \rangle(0)$ & $\langle H_3 \rangle(1)$ & $\langle H_3 \rangle(2)$ & $\langle H_3 \rangle(3)$ & $\langle H_3 \rangle(4)$ \\
\hline
$0.858$ & $0.958$ & $0.758$ & $-2.220$ & $ -1.586$ & $ -0.120$ & $ 1.559$ & $ 2.560$ \\
\hline
$0.420$ & $0.958$ & $0.758$ & $-1.625$ & $-1.081$ &	$0.587$ & $2.224$ & $3.090$\\
$0.550$ & $0.958$ & $0.758$ & $-1.851$ & $-1.042$ &	$-0.184$ & $1.364$ & $2.715$\\
$1.140$ & $0.958$ & $0.758$ & $-1.560$ & $-0.766$ &	$0.128$ & $1.690$ & $3.004$\\
$1.260$ & $0.958$ & $0.758$ & $-1.642$ & $-0.833$ &	$0.293$ & $1.250$ & $3.223$\\
\hline
$0.858$ & $0.190$ & $0.758$ & $-1.323$ & $-0.862$ &	$0.468$ & $2.134$ & $2.590$ \\
$0.858$ & $0.410$ & $0.758$ & $-1.613$ & $-1.034$ & $0.422$ & $1.662$ & $2.877$ \\
$0.858$ & $1.440$ & $0.758$ & $-2.045$ & $-0.700$ & $0.251$ & $1.529$ & $4.683$ \\
$0.858$ & $1.630$ & $0.758$ & $-1.650$ & $-0.659$ & $-0.513$ & $1.819$ & $2.7142$ \\
\hline
$0.858$ & $0.958$ & $-0.510$ & $-2.183$ & $-0.485$ & $1.723$ & $4.450$ & $7.910$ \\
$0.858$ & $0.958$ & $-0.120$ & $-1.392$ & $-0.259$ & $1.564$ & $4.349$ & $5.982$ \\
$0.858$ & $0.958$ & $1.600$  & $-1.523$ & $-0.618$ & $0.598$ & $2.696$ & $4.006$ \\
$0.858$ & $0.958$ & $1.930$  & $-1.755$ & $-0.336$ & $1.475$ & $3.242$ & $6.557$ \\
\hline
\end{tabular}
\caption{Expectation value $\langle H_4 \rangle (r)$ as a function of error parameter $r = 2M + 1$, $M=0,1,2,3$, for various sets of variational parameters that lead to ${\sim}10\%$ or ${\sim}20\%$ deviations from the known energy minimum $\langle H_4 \rangle$. The top row shows the exact minimum configuration and results. The next set of four rows show the cases where we vary $\lambda_0$, while the following two sets of four rows show the corresponding configuration-results pair for varying $\lambda_1$ and $\lambda_2$, respectively.}
\label{tab:basin}
\end{table}
}

\section{Discussion}

In this paper, we showed the quantum computational results obtained from 5- and 7-qubit trapped-ion quantum computers simulating a Deuteron. We improved on the previous result for the three-qubit ansatz and extended the ansatz size beyond the previous state of the art \cite{dumitrescu2018}. Our four-qubit ansatz result of $E_4=-2.220 \pm 0.179$MeV may be compared with the exact Deuteron ground-state energy $-2.224$MeV.

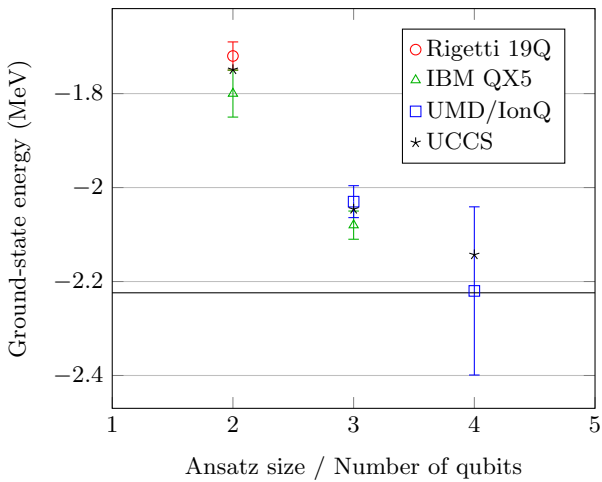
\begin{figure}
\begin{tikzpicture}
    \begin{axis}[
    width=8cm,
    ymajorgrids=true,
    legend style={at={(0.6,0.95)},anchor=north west},
    xlabel={Ansatz size / Number of qubits},
    ylabel={Ground-state energy (MeV)},
    xtick={1,2,3,4,5},
    xmin=1,
    xmax=5,
    every axis legend/.append style={nodes={right}}
    ]
    
\addplot+[only marks, red, mark=o,error bars/.cd, y dir=both,y explicit] coordinates {
         (2,-1.72) +- (1,0.03) 
    };
    \addlegendentry{Rigetti 19Q}

\addplot+[only marks,green!70!black, mark=triangle,error bars/.cd, y dir=both,y explicit] coordinates {
         (2,-1.80) +- (1,0.05)
         (3,-2.08) +- (1,0.03)
    };
    \addlegendentry{IBM QX5}
    
\addplot+[only marks, blue, mark=square,error bars/.cd, y dir=both,y explicit] coordinates {
         (3,-2.030) +- (1,0.034)
         (4,-2.220) +- (1,0.179)
    };
    \addlegendentry{UMD/IonQ}    
\addplot[only marks, black, mark=star] coordinates {
         (2,-1.749)
         (3,-2.046)
         (4,-2.143)	
    };
    \addlegendentry{UCCS}    
\addplot[mark=none, black, samples=2] {-2.224};
  \end{axis}
\end{tikzpicture}
\caption{Aggregate results on the Deuteron simulation performed across different quantum computing platforms. Open symbols denote the experimental results. Star symbols denote the exact UCCS results. The black solid line denotes the exact deuteron ground-state energy.}
\label{fig:comp}
\end{figure}

Figure~\ref{fig:comp} shows the aggregate results, collected from previous studies performed on different quantum computing platforms on the same Deuteron system \cite{dumitrescu2018} and our own results. For the three qubit ansatz, the error margin of the binding energy computed on the IBM QX5 was $3\%$, while it is $0.7\%$ on the IonQ-UMD trapped ion quantum computer at the optimal configuration for the three qubit experiment. Because of the demanding size of the circuit and the susceptibility of NISQ devices to errors, we were unable to run the four-qubit experiments on other quantum computing platforms. We find that, based on Fig.~\ref{fig:comp}, the simulation results converge to the known ground state energy as a function of the ansatz size. We also note that, as expected, the experimental results start deviating more from the exact UCCS results, due to the accumulation of errors. 

Thus, we believe that our EFT simulation may be used as a practical benchmark for quantum computers which characterizes the performance of HQCC algorithms in the presence of noise, alongside the known proposals \cite{QV,QBAS}. We have already successfully implemented the simulation across different platforms (superconducting and trapped-ion quantum computers) and also within the same platform with different configurations (5 and 7 qubit trapped-ion quantum computers). Since our ansatz circuits require only nearest-neighbor connectivity, our benchmark is expected to be readily be implemented across any platform and serve as a baseline, since more complex connectivity available on a quantum computer can only help boost the quantum computational power \cite{Linke3305}. Our HQCC approach will also help benchmark the interface between quantum and classical processors. In this paper, we have taken first steps in this direction. We anticipate using the algorithm to benchmark upcoming quantum information processors.

\comment{
\begin{itemize}
\item Comparison to others (IBM, Rigetti)
\item RE linear? quadratic? (need chi-square test)
\item Why off? Reasons?
\item What next? Tug-of-war?
\item Benchmark aspect
\end{itemize}
}

\comment{
Ideal UCC: $E_2  = -1.749$MeV, $E_3 = -2.046$MeV, $E_4 = -2.143$MeV 

Rigetti: $E_2 = -1.72 \pm 0.03$MeV

IBM: $E_2 = -1.80 \pm 0.05$MeV, $E_3 = -2.08 \pm 0.03$MeV

UMD/IonQ: $E_3 = -2.030 \pm 0.034$MeV, $E_4=-2.220 \pm 0.101$MeV
}
\comment{
Fig ~\ref{fig:plot4q} shows that the four qubit ansatz result deteriorates when we switch from linear to quadratic regression. Finding the best statistical regression technique for extrapolating the result of a VQE algorithm to its zero noise limit is still a very much research problem. We hope to conduct further investigation to develop a rigorous error mitigation protocol.

\begin{figure}[H]
\centering
\includegraphics[scale=0.4]{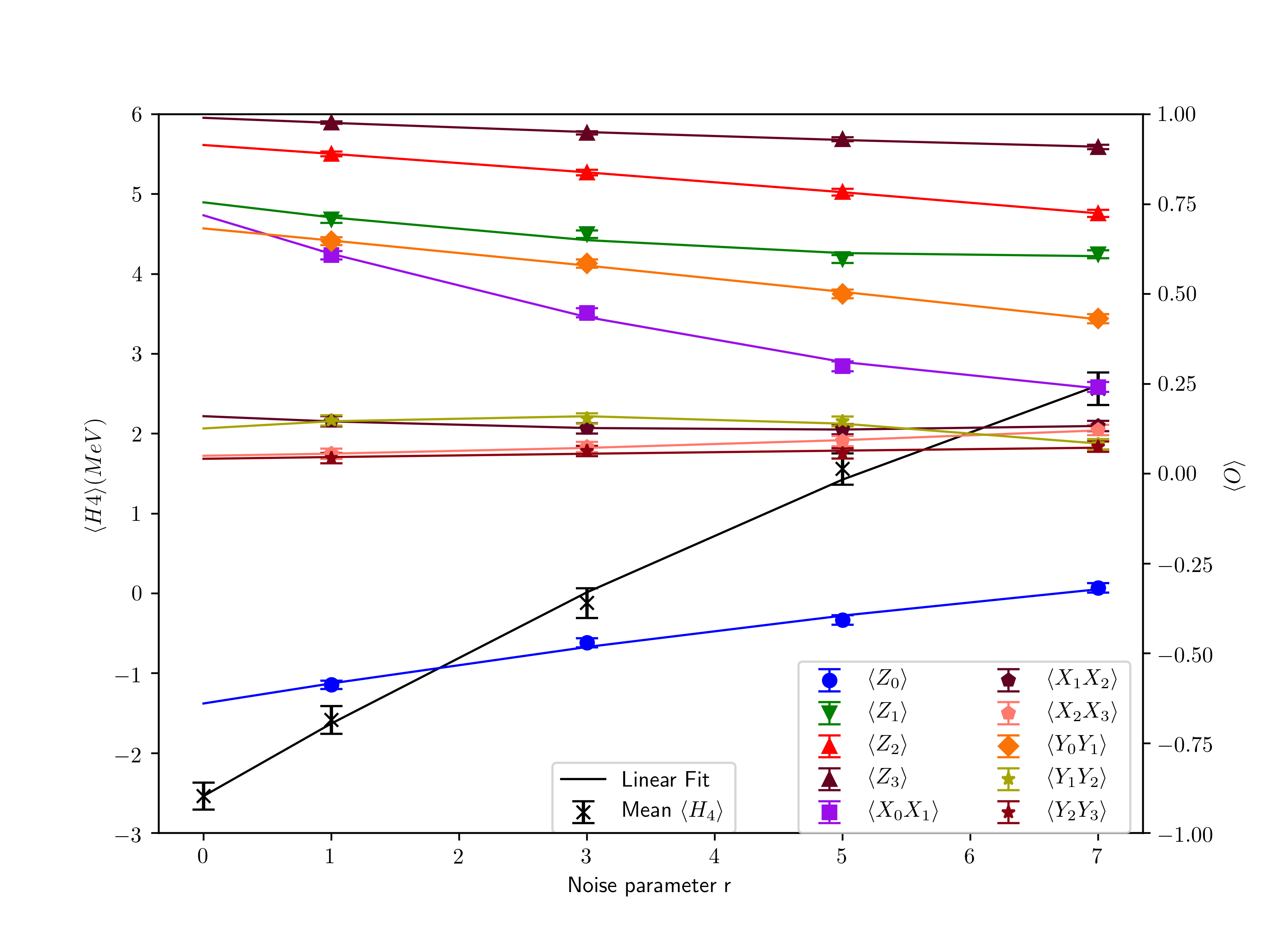}
\caption{Binding energy at the zero noise limit using the four qubit ansatz. Quadratic fit has been used to reach to the zero noise limit. Note that the accuracy has been deteriorated.}
\label{fig:plot4q}
\end{figure}
}

\section{acknowledgments}
This work is supported by the U.S. Department of Energy, Office of Science, Office of Advanced Scientific Computing Research
(ASCR) Quantum Algorithm Teams and Testbed Pathfinder programs, under field work proposal numbers
ERKJ332 and ERKJ335. We thank P.\ Lougovski and E. Dumitrescu for useful discussions. We thank T. Papenbrock for the Hamiltonian and energy extrapolation formula. Some materials presented build upon upon work supported by the U.S. Department of Energy, Office
of Science, Office of Nuclear Physics under Award Nos. DEFG02-96ER40963 and DE-SC0018223 (SciDAC-4 NUCLEI). A portion of this work was performed at Oak Ridge National Laboratory, operated by UT-Battelle for the U.S. Department of Energy under Contract No. DEAC05-00OR22725.


\bibliography{refs}

\section*{supplementary material}

\subsection{Deuteron Hamiltonian}
The deuteron is a shallow bound state of the proton-neutron system with a binding energy of about $B=2.2$~MeV, corresponding to a bound-state momentum $\kappa =\sqrt{2\mu B}\approx 45$~MeV ($\mu$ denotes the reduced mass). This momentum is small compared to other scales such as the pion mass at about 140~MeV, the excitation of the nucleon in a delta-resonance (at about 300~MeV), or the dividing scale  $\Lambda_{\rm QCD}\approx 1$~GeV of quantum chromodynamics (QCD). The ensuing separation of scales allows us to describe the deuteron in pionless  EFT~\cite{vankolck1999,bedaque2002}. As the range of the nuclear interaction is small compared to the inverse bound-state momentum, any short-range central potential can be taken for a leading-order description of the deuteron in pionless EFT. For our purposes, an implementation of the effective field theory directly in the harmonic-oscillator basis~\cite{binder2016}, realized as a discrete variable representation, is convenient. This also allows us to perform infrared extrapolations~\cite{furnstahl2012,coon2012,furnstahl2014} of results obtained in small Hilbert spaces, i.e. employing few qubits, to infinite spaces.  

We consider the deuteron in its center-of-mass system. For the relative coordinate, we choose a harmonic oscillator basis with energy spacing $\hbar\omega=7$~MeV. This yields an oscillator spacing of $b=\sqrt{\hbar/(\mu\omega)}\approx 3.5$~fm. The short-ranged interaction only acts between the $0s$ state, implying an ultraviolet cutoff $\Lambda\approx\sqrt{7}/b\approx 150$~MeV~\cite{konig2014}, and the bound-state momentum fulfills $\kappa \ll \Lambda$ as required for EFT. Thus, the cutoff is close to the breakdown scale (e.g. the pion mass) of pionless EFT.

Ideally one would pick an even larger value for $\Lambda$, either by choosing a larger oscillator spacing or by increasing the number of states where the potential is active. In our case, increasing $\hbar\omega$ further would not yield a bound state (i.e. a ground state with negative energy) when the Hilbert space is limited to a single state. Increasing the number of states where the potential is active would increase the minimum number of qubits required to perform the computation. In this work, we make effort to ensure the calculation is amenable to implementation on existing quantum computers. This motivates our current choice of parameters.

\subsection{Ansatz Circuit for Trapped-Ion Quantum Computer}

In order to apply the circuit that implements the ansatz state defined in Eq.~(\ref{Eq:UCC_hopping}) on a trapped-ion quantum computer, we rewrite the quantum circuit ${\mathcal C}_N$ over the native gate set amenable to implementation on a trapped-ion quantum computer. To do so, we start with useful circuit identities for those gates that appear in ${\mathcal C}_N$, decomposed into trapped-ion quantum computer native gates, as shown below.

\begin{align*}
\label{eq:iden}
\raisebox{-0.01em}{\Qcircuit @C=1em @R=1.5em {
  &\ctrl{1}& \qw\\
 &\targ&\qw
}} 
&\raisebox{-1.15em}{\hspace{2mm}=\hspace{4mm}}
\scalebox{0.85}{\Qcircuit @C=1em @R=.7em {
 & \gate{Y_{\frac{\pi}{2}}} & \multigate{1}{XX_{\frac{\pi}{2}}}& \gate{X_{-\frac{\pi}{2}}}& \gate{Y_{-\frac{\pi}{2}}}&  \qw\\
 & \qw&\ghost{XX_{\frac{\pi}{2}}}&\gate{X_{-\frac{\pi}{2}}}&\qw&\qw
}}
\nonumber\\
\Qcircuit @C=1em @R=.9em {
 & \ctrl{1}& \qw\\
 & \gate{Y_{\theta}}&\qw
} 
&\raisebox{-1.15em}{\hspace{2mm}=\hspace{4mm}}
\Qcircuit @C=1em @R=.7em {
 & \qw& \ctrl{1}& \qw& \ctrl{1}& \qw\\
 & \gate{Y_{\frac{\theta}{2}}}&\targ&\gate{Y_{-\frac{\theta}{2}}}&\targ&\qw
}
\nonumber\\
\raisebox{-0.01em}{\Qcircuit @C=1em @R=.7em {
 & \ctrl{1}&\qw &\ctrl{1}& \qw\\
 & \targ& \gate{Y_{\theta}}&\targ&\qw
}} 
&\raisebox{-1.15em}{\hspace{2mm}=\hspace{4mm}}
\scalebox{0.65}{\Qcircuit @C=1em @R=.7em {
 & \gate{X_{-\frac{\pi}{2}}}&\gate{Z_{-\frac{\pi}{2}}} & \multigate{1}{XX_{\theta}}&\gate{Z_{\frac{\pi}{2}}}&\gate{X_{\frac{\pi}{2}}}&  \qw\\
 & \qw&\gate{Z_{-\frac{\pi}{2}}}&\ghost{XX_{\theta}}&\gate{Z_{\frac{\pi}{2}}}&\qw
}}
\end{align*}

Using the identities \cite{maslov2017basic, maslov2018use}, we obtained the ansatz-preparation circuits that are amenable to implementation on a trapped-ion quantum computer. We then optimized these circuits using known rules (see for instance \cite{maslov2018use}), reducing the number of XX gates and RX gates, at the cost of, e.g., increasing the number of RZ gates. We chose to do so since on our quantum computer it is more costly to implement XX and RX gates than RZ gates. Figure ~\ref{fig:opt4} shows an exemplary case of ${\mathcal C}_4$.

\begin{widetext}
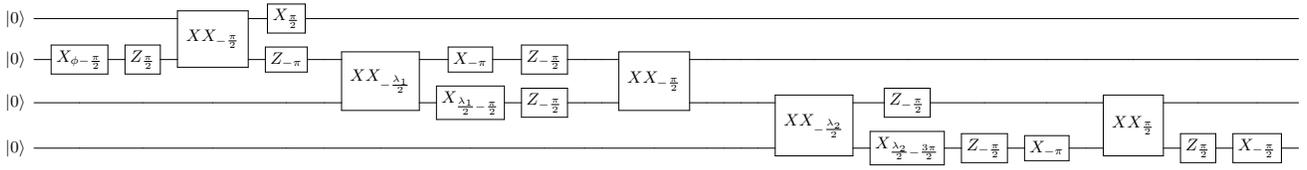
\begin{figure*}[t]
\centering
\scalebox{0.7}{
\Qcircuit @C=1em @R=.7em {
\lstick{|0\rangle}  &\qw &\qw&\multigate{1}{XX_{-\frac{\pi}{2}}} &\gate{X_{\frac{\pi}{2}}}&\qw &\qw&\qw&\qw&\qw& \qw&\qw &\qw&\qw&\qw&\qw&\qw& \qw & \qw& \qw& \qw&\qw&\qw&\qw&\qw\\
\lstick{|0\rangle} &\gate{X_{\phi - \frac{\pi}{2}}} &\gate{Z_{\frac{\pi}{2}}}&\ghost{XX_{-\frac{\pi}{2}}}&\gate{Z_{-\pi}}&\qw& \multigate{1}{XX_{-\frac{\lambda_1}{2}}}&\gate{X_{-\pi}}&\gate{Z_{-\frac{\pi}{2}}}&\qw&  \qw&\multigate{1}{XX_{-\frac{\pi}{2}}}&\qw&\qw&\qw& \qw &\qw&\qw&\qw&\qw&\qw&\qw&\qw&\qw&\qw\\
\lstick{|0\rangle}&\qw&\qw&\qw&\qw&\qw&\ghost{XX_{-\frac{\lambda_1}{2}}}&\gate{X_{\frac{\lambda_1}{2}-\frac{\pi}{2}}}&\gate{Z_{-\frac{\pi}{2}}}&\qw&\qw&\ghost{XX_{-\frac{\pi}{2}}}&\qw&\qw&\qw&\qw&\multigate{1}{XX_{-\frac{\lambda_2}{2}}}&\gate{Z_{-\frac{\pi}{2}}}&\qw&\qw&\qw&\multigate{1}{XX_{\frac{\pi}{2}}}&\qw &\qw & \qw\\
\lstick{|0\rangle} &\qw&\qw&\qw&\qw&\qw&\qw&\qw&\qw&\qw&\qw&\qw&\qw&\qw&\qw&\qw&\ghost{XX_{-\frac{\lambda_2}{2}}}&\gate{X_{\frac{\lambda_2}{2}-\frac{3 \pi}{2}}}&\gate{Z_{-\frac{\pi}{2}}}&\gate{X_{-\pi}}&\qw&\ghost{XX_{\frac{\pi}{2}}}&\gate{Z_{\frac{\pi}{2}}}&\gate{X_{-\frac{\pi}{2}}}& \qw 
}}
\caption{Optimized four qubit ansatz circuit ${\mathcal C}_4$, written over a native gate set for trapped-ion quantum computers.} \label{fig:opt4}
\end{figure*}
\end{widetext}

\end{document}